\documentclass[journal,10pt]{IEEEtran}
\usepackage[T1]{fontenc}
\usepackage{amsmath,amssymb,amsfonts,bm,bbm,stmaryrd,tikz,pgfplots}
\usetikzlibrary{calc,shapes,decorations.pathreplacing,chains,matrix,arrows,intersections}
\usepackage{amsbsy}
\usepackage{graphicx}
\usepackage{graphics}
\usepackage{epstopdf}
\usepackage{algorithm}
\usepackage{algorithmic}
\usepackage{subfigure}
\usepackage{cite}
\usepackage{slashbox}
\usepackage{multirow}
\usepackage{ctable}
\usepackage{flushend}
\usepackage{etex}
\usepackage{hyperref}
\renewcommand{\sc}{spatially-coupled}
\global\long\def\dl{\texttt{l}}
\global\long\def\dr{\texttt{r}}
\newcommand{\rate}{\mathtt{R}}

\renewcommand{\a}{\mathsf{a}}

\newcommand{\y}{\bm{y}}
\newcommand{\x}{\mathsf{x}}
\renewcommand{\u}{\bm{u}}
\newcommand{\X}{\bm{X}}
\newcommand{\Xc}{\mathcal{X}}
\newcommand{\Xs}{\mathcal{X}}

\providecommand{\abs}[1]{\left\lvert#1\right\rvert}
\newcommand{\Y}{\mathbf{Y}}

\global\long\def\amap{\alpha^{\mathrm{MAP}}}
\global\long\def\gbp{\mathsf{g}^{\mathrm{BP}}}
\global\long\def\gmap{\mathsf{g}}
\global\long\def\gfun{\mathsf{G}}
\global\long\def\amapub{\bar{\alpha}}

\newtheorem{theorem}{{Theorem}}
\newtheorem{lemma}[theorem]{{Lemma}}
\begin{document}
\pgfdeclarelayer{background}
\pgfdeclarelayer{foreground}
\pgfsetlayers{background,main,foreground}
\title{Performance of Spatially-Coupled LDPC Codes and Threshold Saturation over BICM Channels}

\author{Arvind~Yedla,~\IEEEmembership{Member,~IEEE,}
        Mostafa~El-Khamy,~\IEEEmembership{Senior Member,~IEEE,}
        Jungwon~Lee,~\IEEEmembership{Senior Member,~IEEE,}
        and~Inyup~Kang,~\IEEEmembership{Member,~IEEE}}% \\% <-this % stops a space
\maketitle

\begin{abstract}
%\boldmath
  We study the performance of binary spatially-coupled low-density
  parity-check codes (SC-LDPC) when used with bit-interleaved
  coded-modulation (BICM) schemes. This paper considers the cases when
  transmission takes place over additive white Gaussian noise (AWGN)
  channels and Rayleigh fast-fading channels. The technique of upper
  bounding the maximum-a-posteriori (MAP) decoding performance of LDPC
  codes using an area theorem is extended for BICM schemes. The upper
  bound is computed for both the optimal MAP demapper and the
  suboptimal max-log-MAP (MLM) demapper. It is observed that this
  bound approaches the noise threshold of BICM channels for regular
  LDPC codes with large degrees. The rest of the paper extends these
  techniques to SC-LDPC codes and the phenomenon of threshold
  saturation is demonstrated numerically. Based on numerical evidence,
  we conjecture that the belief-propagation (BP) decoding threshold of
  SC-LDPC codes approaches the MAP decoding threshold of the
  underlying LDPC ensemble on BICM channels. Numerical results also
  show that SC-LDPC codes approach the BICM capacity over different
  channels and modulation schemes.
\end{abstract}


% \fi
\IEEEpeerreviewmaketitle

\section{Introduction}
\label{sec:introduction}
%\IEEEPARstart{T}{he} 
The phenomenon of threshold saturation was introduced
by Kudekar et al. \cite{Kudekar-it11} to explain the impressive
performance of convolutional low-density parity-check (LDPC) ensembles
\cite{Felstrom-it99,Lentmaier-isit05}. These codes are essentially
terminated convolutional codes with large memory, which admit a sparse
parity-check matrix representation. One way to construct these codes
is to ``spatially-couple'' an underlying LDPC ensemble, resulting in a
\sc{} LDPC (SC-LDPC) ensemble. It was observed that the
belief-propagation (BP) threshold of a spatially-coupled ensemble is
very close to the maximum-a-posteriori (MAP) threshold of its
underlying ensemble; a similar statement was formulated independently,
as a conjecture in \cite{Lentmaier-isit10}. This phenomenon has since
been called ``threshold saturation via spatial coupling''. Kudekar et
al. prove in \cite{Kudekar-it11} that threshold saturation occurs for
the binary erasure channel (BEC) and a particular class of underlying
regular LDPC ensembles. For general binary-input memoryless symmetric
(BMS) channels, threshold saturation was empirically observed
first~\cite{Lentmaier-it10,Kudekar-istc10} and then analytically shown
~\cite{Kudekar-isit12,Kumar-aller12}. It is known that the MAP
threshold of regular LDPC codes approaches the Shannon limit for
binary memoryless symmetric (BMS) channels with increasing left
degree, while keeping the rate fixed (though such codes have a
vanishing BP threshold) \cite{Kudekar-it11}. So, spatial coupling
provides us with a technique to construct a single capacity
approaching code ensemble for all BMS channels with a given
capacity. This technique is indeed very general and has since been
applied to a broad class of graphical models. A good summary of recent
applications of spatial coupling can be found in~\cite{Kumar-aller12}.

In this paper, we evaluate the performance of \sc{} LDPC codes using
BICM schemes for transmission over additive white Gaussian noise
(AWGN) and Rayleigh fast-fading channels. The noise threshold, a.k.a.
the Shannon limit, for bit-interleaved coded-modulation (BICM) schemes
can be computed using Monte-Carlo simulations via the generalized
mutual information (GMI)~\cite{Nguyen-tcom11}. This method can be used
to compute the information theoretic limits for different suboptimal
BICM schemes and is briefly reviewed in Section~\ref{sec:bicm_cap}. We
review density evolution (DE) for BICM schemes, described
in~\cite[Sec. 5.2]{Caire-2008}, in Section~\ref{sec:bicm_de}. We note
that the above DE can be greatly simplified by using the Gaussian
mixture approximation for the BICM bit-channels presented
in~\cite{Alvarado-tcom09}, to obtain approximate
thresholds. Section~\ref{sec:gexit-curves} extends the GEXIT analysis
and the upper bounding technique on the MAP decoding threshold for
BICM schemes. Section~\ref{sec:spat-coupl-ldpc} extends the analysis
to SC-LDPC codes. The DE results of SC-LDPC codes are presented in
Section~\ref{sec:results} and some concluding remarks are given in Section~\ref{sec:concluding-remarks}.

%%% Local Variables: 
%%% mode: latex
%%% TeX-master: "bicm_paper"
%%% End: 

\section{Background}

\subsection{The BICM Model}
\label{sec:bicm_model}
BICM is a practical approach to coded modulation and was introduced by
Zehavi in~\cite{Zehavi-tcom92}. A comprehensive analysis for BICM is
provided in~\cite{Caire-2008}, which is an excellent reference for
BICM. We now briefly describe the BICM model and the problem
setup. Consider transmission over a memory-less channel with input
alphabet $\Xs$ (with $\abs{\Xs} = 2^M,M\in\mathbb{N}$) and output
alphabet $\mathbb{C}$. We use uppercase letters (e.g., $X, Y$) to
denote random variables and lowercase letters (e.g. $x, y$) to denote
their corresponding realizations. The channel output is given by
\begin{align}
\label{eq:chan}
  Y = AX + Z,
\end{align}
where $X\in\Xs$, $Y\in\mathbb{C}$, and $Z$ is additive Gaussian noise
with variance $\sigma^2$ i.e., $Z\sim\mathcal{CN}(0,\sigma^2)$. We
consider the cases of no fading ($A=1$) and Rayleigh fast-fading
($A\sim\mathcal{CN}(0,1)$). Furthermore, we assume that the receiver
has perfect channel state information for simplicity. The analysis can
be easily extended to the case when the receiver does not have access
to the channel state information~\cite[Sec. 5.1]{Hou-jsac01,RU-2008}.

\begin{figure}
  \centering
  \begin{tikzpicture}[>=stealth,xscale=0.63,
                    mynode/.style={
                      rectangle,
                      minimum size=6mm,
                      very thick,
                      draw=black!50,
                      font=\ttfamily,
                      transform shape}]
  \node (sou) [mynode]               {Src.};
  \node (enc) [mynode,right=of sou] {Enc.};
  \node (map) [mynode,right=of enc] {Mapper};
  \node (chn) [mynode,right=of map] {Channel};
  \node (dem) [mynode,right=of chn] {Demapper};
  \node (dec) [mynode,right=of dem] {Dec.};
  \path[thick] (sou) edge[->] (enc)
        (enc) edge[->] (map)
        (map) edge[->] node[above,midway] {$X$} (chn)
        (chn) edge[->] node[above,midway] {$Y$} (dem)
        (dem) edge[->] (dec);
  \draw[thick,dashed,->] 
  % start slightly below center
  ($(dec.east)-(0,2mm)$) 
  % go right
  -- ++(0.25,0) 
  % go down
  -- ++(0,-0.5)
  % go left
  -- ($(dem.west)-(0.45,2mm)-(0,0.5)$)
  % back to demapper
  |- ($(dem.west)-(0,2mm)$);
\end{tikzpicture}
%%% Local Variables: 
%%% mode: latex
%%% TeX-master: "../bicm_paper"
%%% End: 
  \caption{The BICM system model. The codeword is mapped to a symbol
    $X\in\Xs$ using the mapper. The channel output $Y$ is then passed
    through a demapper which performs the symbol-to-bit metric
    calculation. The decoder output can be optionally fed back to the
    demapper as apriori information for BICM-ID.}
  \label{fig:sys-model}
\end{figure}
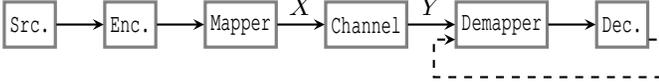

A Bernoulli-($1/2$) source is encoded using an LDPC code chosen
uniformly at random from the standard irregular ensemble
LDPC$(N,\lambda,\rho)$~\cite[Ch. 3]{RU-2008}. Here, $\lambda(x) =
\sum_i \lambda_i x^{i-1}$ is the degree distribution (from an edge
perspective) corresponding to the variable nodes and $\rho(x) = \sum_i
\rho_i x^{i-1}$ is the degree distribution (from an edge perspective)
of the parity-check nodes in the decoding graph.\footnote{The edges of
  the variable nodes connected to the demapper are not included in the
  degree profile.} The coefficient $\lambda_i$ (resp. $\rho_i$) gives
the fraction of edges that connect to variable nodes
(resp. parity-check nodes) of degree $i$. Likewise, let $L_i$ be the
fraction of variable nodes with degree $i$ and define $L(x) = \sum_i
L_ix^i$. The design rate of the LDPC code is given by
\begin{align*}
  \rate(\lambda,\rho) = 1 -
  \frac{\int_0^1\rho(x)\text{d}x}{\int_0^1\lambda(x)\text{d}x}.
\end{align*}
The blocklength $N$ is assumed to be a multiple of $M$, where groups
of $M$ bits are mapped to a symbol in $\Xs$ and then transmitted over
the channel. At the receiver, a demapper first performs the
symbol-to-bit metric calculation based on the received symbol $Y$, and
the metrics are then passed to the decoder. One can also perform the
symbol-to-bit metric calculation iteratively, by using the decoder
output as apriori information at the demapper. This scheme is commonly
known in the literature as BICM iterative detection (BICM-ID). The
block diagram of a general BICM system is shown in
Fig.~\ref{fig:sys-model}.

A BICM scheme is specified by the bit-to-symbol mapper and the
demapper. Throughout this work, we consider square quadrature
amplitude modulation (QAM) constellations with the Gray mapping
scheme, and the optimal MAP and suboptimal max-log-MAP (MLM)
demappers.

%%% Local Variables: 
%%% mode: latex
%%% TeX-master: "bicm_paper"
%%% End: 

\subsection{Noise Threshold of BICM Channels}
\label{sec:bicm_cap}
Consider the case when the demapper calculation is not updated between
iterations. The performance of the BICM scheme (for optimal demappers)
is given by the capacity of a set of parallel independent
channels~\cite{Caire-it98}. It was also characterized in terms of the
generalized mutual information (GMI) by viewing the BICM decoder as a
mismatched decoder~\cite{Martinez-it09}. The GMI analysis was used
in~\cite{Jalden-itw10} to compute the performance of BICM in the
presence of suboptimal demappers. The achievable information rate of a
given BICM scheme can be computed using Monte-Carlo simulations via
the GMI~\cite{Nguyen-tcom11}. Consider a BICM channel with $M$-bits
per symbol. The $I$-curve was introduced in~\cite{Nguyen-tcom11} and
can be computed for the $m$-th bit level via
\begin{align*}
 % \label{eq:i_bicm}
  I_m(s;\sigma)\! =\! 1\! - \mathbb{E}_{X,Y}\!\log\bigl(1\! +\! \text{exp}\bigl((2b_m(X)\!-\!1)\Lambda_m(Y(\sigma))s\bigr)\bigr),
\end{align*}
where $X$ and $Y$ are the channel input and output respectively (we
have used the notation $Y(\sigma)$ to make explicit the dependence on
the channel noise variance), $b_m(X)$ is the $m$-th bit label of
symbol $X$ and $\Lambda_m(Y)$ is the log-likelihood ratio of the
$m$-th bit of the symbol after passing through the
demapper~\cite{Nguyen-tcom11}. The $I$-curve of the BICM channel is
then computed as
\begin{align*}
  I(s;\sigma) = \sum_{m=0}^{M-1}I_m(s;\sigma).
\end{align*}
The achievable information rate of the BICM scheme is equal to the
GMI, given by $I(\sigma) = \max_s I(s;\sigma)$. This enables us to
compute the achievable information rate of BICM schemes, for different
modulation schemes, bit-to-symbol mappings and demappers. The noise
threshold for error-free transmission at a given transmission rate $R$
can then be computed for a specific BICM scheme by $\sigma^* =
I^{-1}(R)$.

For the case of BICM-ID, the capacity would be equal to the coded
modulation capacity, when the input alphabet is restricted to $\Xs$.
The noise threshold can be computed for this case similarly. Let
$I(\sigma) = I(X;Y(\sigma))$, when the input is uniformly
distributed. Then the noise threshold for BICM-ID is given by
$\sigma^* = I^{-1}(R)$.

%%% Local Variables: 
%%% mode: latex
%%% TeX-master: "bicm_paper"
%%% End: 

\subsection{Density Evolution}
\label{sec:bicm_de}
We begin this section by first introducing some notation. Let $v_i$,
$c_j$ and $d_k$ denote the variable, check and demapper nodes
respectively. Let $\pi(k,m)\triangleq (k-1)M+m = i$, be the mapping
from the demapper nodes to the variable nodes i.e., the $m$-th bit of
demapper node $k$ is connected to variable node $i$. When the symbol
index $k$ is understood from context, we write $\pi(m) = i$, and
$m=\pi^{-1}(i)$. The $m$-th bit corresponding to $x\in\Xs$ is denoted
by $b_m(x)$, and the set of symbols where the $m$-th bit is zero (one)
is denoted by $\Xs_0^m$ ($\Xs_1^m$).

\begin{figure}
  \centering
  \begin{tikzpicture}[scale=0.53,>=stealth]

\draw[rounded corners,fill=brown,opacity=0.4] (-1,2) rectangle +(13,1);
\draw (5.5,2.5) node {\ttfamily{permutation} $\pi$};

\foreach \x in {0,2,4,6,11}
{
  % Bottom Check Nodes
  \filldraw[black] (\x,0.9)+(-4pt,-4pt) rectangle +(4pt,4pt);
  % Bottom Check Edges
  \draw (\x,0.9)+(0pt,-4pt) -- (\x,2);
  \draw (\x,0.9)+(0pt,-4pt) -- ([xshift = 3pt]\x,2);
  \draw (\x,0.9)+(0pt,-4pt) -- ([xshift = 6pt]\x,2);
  \draw (\x,0.9)+(0pt,-4pt) -- ([xshift = -3pt]\x,2);
  \draw (\x,0.9)+(0pt,-4pt) -- ([xshift = -6pt]\x,2);
}

\foreach \x in {-0.5,1.5,3.5,5.5,9.5,11.5} {
  % Bottom Variable Nodes
  \shade[ball color=blue] (\x,4) circle (4pt);
  % Bottom Variable Edges
  \begin{pgfonlayer}{background}
    \draw (\x,4)+(0cm,-2pt) -- ([xshift=0cm]\x,3);
    \draw (\x,4)+(0cm,-2pt) -- ([xshift=-0.2cm]\x,3);
    \draw (\x,4)+(0cm,-2pt) -- ([xshift=0.2cm]\x,3);
  \end{pgfonlayer}
}

% Correlation Nodes
\foreach \x in {0.5,4.5,10.5} {
  \node[diamond,draw=black,fill=red,inner sep=0pt,minimum size=6pt] at (\x,5)
  {};
  \begin{pgfonlayer}{background}
    \draw[->] (\x,5)+(0,1) -- ([yshift=5pt]\x,5);
  \end{pgfonlayer}
}

% Correlation Node Connections
\begin{pgfonlayer}{background}
\draw (-1,4)+(0.5cm,2pt) -- (0.5,5);
\draw (1,4)+(0.5cm,2pt) -- (0.5,5);
\draw (3,4)+(0.5cm,2pt) -- (4.5,5);
\draw (5,4)+(0.5cm,2pt) -- (4.5,5);
\draw (9,4)+(0.5cm,2pt) -- (10.5,5);
\draw (11,4)+(0.5cm,2pt) -- (10.5,5);
\end{pgfonlayer}

% Gap between nodes
\foreach \x in {6.5,7,7.5,8,8.5} {
  \foreach \y in {4,5} {
    \filldraw (\x,\y) circle (1pt);
  }
}
\foreach \x in {7.5,8,8.5,...,9.5} 
    \filldraw (\x,1) circle (1pt);
\foreach \x in {0.25,0.5,0.75,4.25,4.5,4.75,10.25,10.5,10.75} 
    \filldraw (\x,3.5) circle (0.5pt);

\draw (-2,1.5) node {$\rho(x)$};
\draw (-2,3.5) node {$\lambda(x)$};

\draw (12.7,5) node[black] {$\phi(\cdot;\sigma)$};

\draw[->,thick,gray] (12.5,3.5) -- (12.5,1.5)  node[black,midway,right=2pt] {$\mathsf{a}^{(\ell)}$};
\end{tikzpicture}

%%% Local Variables: 
%%% mode: latex
%%% TeX-master: "../bicm_paper"
%%% End: 
  \caption{The Tanner graph at the decoder for BICM channels. The red
    diamonds represent the demapper nodes, the blue circles and black
    squares represent the variable and check nodes respectively. Each
    demapper node is connected to $M$ variable nodes.}
  \label{fig:bicm_tanner}
\end{figure}
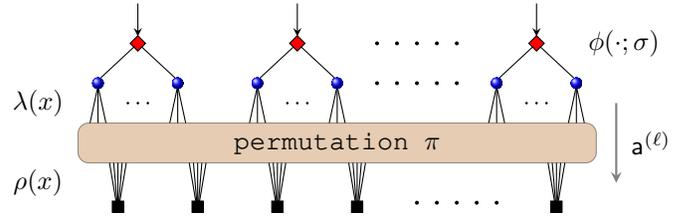

The factor graph structure at the joint decoder is shown in
Fig.~\ref{fig:bicm_tanner}. The joint decoder proceeds by performing
one round of decoding for the LDPC code followed by a demapper
update. This is the schedule for BICM-ID. To reduce the complexity
non-iterative detection is used, where the demapper update is not
performed between iterations.

Let $\mu_{v_i\to c_j}^{(\ell)}$, $\mu_{c_j\to v_i}^{(\ell)}$,
$\mu_{v_i\to d_k}^{(\ell)}$ and $\mu_{d_k\to v_i}^{(\ell)}$ be the
messages from the bit node to check node, check node to variable node,
variable node to demapper node and demapper node to variable node
during iteration $\ell$ respectively. All the messages are in
log-likelihood ratio domain. The message passing rules at the variable
and check nodes are the standard rules and their description is
omitted. Using the notation $\partial i$ to denote the set of check
nodes connected to variable node $i$, the message $\mu_{v_i\to
  d_k}^{(\ell)}$ is given by
\begin{align*}
  \mu_{v_i\to d_k}^{(\ell)} = \sum_{j\in \partial i}\mu_{c_j\to v_i}^{(\ell)}.
\end{align*}
Let $m$ be the bit index corresponding to variable node $v_i$ i.e.,
$m=\pi^{-1}(i)$ and $B_m$ be the random variable corresponding to that
bit. The bit probabilities in the $\ell$-th iteration can be computed
using the variable node to demapper node messages via
\begin{align*}
{\Pr}(B_m \!=\! 0) = \frac{\text{e}^{\mu_{v_i\to d_k}^{(\ell)}}}{1 + \text{e}^{\mu_{v_i\to d_k}^{(\ell)}}},\, {\Pr}(B_m \!=\! 1) = \frac{1}{1 + \text{e}^{\mu_{v_i\to d_k}^{(\ell)}}}.
\end{align*}
So, the demapper to variable node message is given by
\begin{align}
\label{eq:demap_update}
 \mu_{d_k\to v_i}^{(\ell)} \!=
 \log\frac{\sum_{x\in\mathcal{X}_0^{m}}p(y_k|x)\prod_{l\neq m}{\Pr}(B_{\pi(l)} \!=\! b_{\pi(l)}(x))}{\sum_{x\in\mathcal{X}_1^{m}}p(y_k|x)\prod_{l\neq m}{\Pr}(B_{\pi(l)} \!=\! b_{\pi(l)}(x))},
\end{align}
where $B_{\pi(l)}$ is used to denote the bit corresponding to variable
node $v_{\pi(k,l)}$ and is an abuse of notation. The above message
passing rule is for the optimal MAP demapper. The rule for the MLM
demapper is obtained by performing the standard approximation of the
above equation to reduce complexity. The variable nodes $i=1,\cdots,N$
can be grouped into equivalence classes via the function $\pi(k,m)$
i.e., let $V_m=\{v_i|\pi(k,m)=i,k=1,\cdots,N/M\}$ denote the set of
all variable nodes connected to the $m$-th bit of the demapper
nodes. Denote the density of messages emanating from the variable
nodes in $V_m$ to the check nodes at iteration $\ell$ by
$\a_m^{(\ell)}$, conditioned on the transmission of an all-zero
codeword. Note that the all-zero codeword assumption is not valid, but
we can still use DE with standard symmetrizing
techniques~\cite[Ch. 7]{Hou-it03,RU-2008}. The transformation of the
densities of the incoming messages at the check node and variable node
are denoted by $\boxast$ and $\varoast$ respectively (see discussion
in \cite[p.  181]{RU-2008}). For a density $\x$, we denote
\begin{align*}
 \x^{\boxast n} \triangleq \underbrace{\x\boxast\x\boxast\cdots\boxast\x}_n,
\end{align*}
and likewise for $\x^{\varoast n}$. Using this notation, define
$\lambda(\x)=\sum_i\lambda_i\x^{\varoast(i-1)}$,
$\rho(\x)=\sum_i\rho_i\x^{\boxast(i-1)}$ and $L(\x) = \sum_i L_i\x^i$.

The density of messages at the input to the check nodes is given by  
\begin{align}
\label{eq:avg_density}
  \a^{(\ell)} = \frac 1M \sum_{m=1}^M\a_m^{(\ell)}.
\end{align}
We call $\a^{(\ell)}$ the average density of messages from the
variable node to check node at iteration $\ell$. The density of
messages from the variable node to the demapper node is then given by
$L(\rho(\a^{(\ell)}))$. Let $\phi_m(\cdot;\sigma)$ be the demapper
density transformation operator of the $m$-th bit corresponding to
\eqref{eq:demap_update}. Then, the density evolution equations are
given by
\begin{align}
\label{eq:bicm-de}
  \a_m^{(\ell+1)} &= \phi_m\left(L(\rho(\a^{(\ell)}));\sigma\right)\varoast\lambda\left(\rho(\a^{(\ell)})\right)\notag\\
  \a^{(\ell)} &= \frac 1M \sum_{m=1}^M\a_m^{(\ell)},
\end{align}
from which one obtains the recursion
\begin{align*}
  \a^{(\ell+1)} %&= \frac 1m \sum\left(\phi_l\left(L(\rho(\b^{(\ell)}));\sigma\right)\varoast\lambda\left(\rho(\b^{(\ell)})\right)\right)\\
%  &= \left(\frac 1m \sum_{l=1}^m \phi_l\left(L(\rho(\b^{(\ell)}));\sigma\right)\right)\varoast\lambda\left(\rho(\b^{(\ell)})\right)\\
&= \phi\left(L(\rho(\a^{(\ell)}));\sigma\right)\varoast\lambda\left(\rho(\a^{(\ell)})\right),
\end{align*}
where $\phi(\cdot;\sigma)$ maps the incoming density at the demapper
node to the average output density i.e., 
\begin{align*}
  \phi(\mathsf{x};\sigma) = \frac 1M \sum_{m=1}^M\phi_m(\mathsf{x};\sigma).
\end{align*}
One can use the `$M$' equations \eqref{eq:bicm-de} to perform DE for
protograph based LDPC codes to design bit mappings for optimal
performance, similar to~\cite{Nguyen-globe11} where the authors use
PEXIT curves for optimization. This function does not have a closed
form expression and can be computed using Monte-Carlo simulations.

%%% Local Variables: 
%%% mode: latex
%%% TeX-master: "bicm_paper"
%%% End: 

%\IEEEpubidadjcol

\section{GEXIT Curves for BICM}
\label{sec:gexit-curves}
In this section, we derive an expression for the BP-GEXIT curve for
LDPC codes for BICM schemes. Using the BP-GEXIT curve and the area
theorem, an upper bound is derived for the MAP decoding threshold of
LDPC codes for BICM schemes. As defined in the previous section, let
$\pi(k,m)= (k-1)M+m = i$, be the mapping from the demapper nodes to
the variable nodes. In this section, boldface uppercase letters
(e.g. $\X,\Y$) are used to denote random vectors and $\X_{\sim k}$ to
denote the vector with all elements of $\X$ except the $k$-th element.
Let $x_k^{[m]}$ be the $m$-th bit of symbol $k$ and let $i =
\pi(k,m)$. Throughout this section, variable node $v_i$ shall be
denoted by $x_k^{[m]}$ via the function $\pi$. Consider transmission
over the BICM channel family~\eqref{eq:chan} parametrized by the
normalized channel entropy per bit, given by
\begin{align*}
  \alpha = \frac 1M H(X|Y).
\end{align*}
We note that $\alpha\in[0,1]$ and that the channel family is complete
and degraded with respect to $\alpha$. % So, one can define a threshold
% on the channel parameter $\alpha$ for succesfull decoding.
Let $\X \in\Xs^{N/M}$ be the transmitted vector and $\Y$ be the output
of the channel. Following the definition for BMS
channels~\cite[Ch. 4]{RU-2008}, the GEXIT function for BICM channels
is defined as

\begin{align*}
\gmap(\alpha) = \frac 1N \frac{\text{d}H(\X|\Y(\alpha))}{\text{d}\alpha},
\end{align*}
and satisfies an area theorem by definition:

\begin{align}
  \label{eq:area_bicm}
  \int_{0}^{1}\gmap(\alpha)\text{d}\alpha &= \frac 1N\left(H(\X|\Y(1)) - H(\X|\Y(0))\right) \notag\\
  &= \rate(\lambda,\rho).
\end{align}

It is convenient to assume that symbol $k$ is transmitted through a
channel with parameter $\alpha_k$, and that each $\alpha_k$ is further
characterized by a common parameter $\alpha$ in a smooth and
differentiable manner. For the case under consideration, we simply
have $\alpha_k = \alpha$. Then define the $k$-th GEXIT function
\begin{align*}
\gmap_k(\alpha_1,\cdots,\alpha_{N/M}) = \frac{\partial H(\X|\Y(\alpha_1,\cdots,\alpha_{N/M}))}{\partial\alpha_k}. 
\end{align*}
So, the GEXIT function is given by
\begin{align*}
  \gmap(\alpha) = \frac 1N \sum_{k=1}^{N/M}\gmap_k(\alpha_1,\cdots,\alpha_{N/M})\frac{\partial \alpha_k}{\partial \alpha}.
\end{align*}
\begin{lemma}
  Consider transmission using an LDPC code from the ensemble
  LDPC$(N,\lambda,\rho)$ over the BICM channel with parameter
  $\alpha$. Define $\phi_k(\y_{\sim k}) = \{p_{X_k|\Y_{\sim
      k}}(x|\y_{\sim k}), x\in \Xc\}$ and let $\Phi_k(\Y_{\sim k})$ be
  the corresponding random variable. Then, the $k$-th GEXIT function is given
  by
  \begin{align}
    \label{eq:lgexit}
    \gmap_k(\alpha) = \gfun\left(\x_k(\u);\alpha\right) \triangleq\sum_{x_k\in\Xc}p(x_k)\int_{\u}\x_{x_k}(\u)\kappa_{x_k}(\u)\text{d}\u,
  \end{align}
  where $\x_k(\u) = \{\x_{x_k}(\u),x_k\in\Xs\}$, and $\x_{x_k}(\u) = p(\phi_k|x_k)$ is the distribution of $\phi_k$
  assuming that $X_k=x_k$ was transmitted, and the GEXIT kernel is given by
  \begin{align}
    \label{eq:kernel_bicm}
    \kappa_x(\u) =
    \int_y\frac{\partial}{\partial\alpha}p(y|x)\log_2\frac{\sum_{x'\in\Xc}\u[x']p(y|x')}{\u[x]p(y|x)}\text{d}y,
  \end{align}
  where $\u[j]$ denotes the $j$-th component of $\u$.
\end{lemma}
\begin{IEEEproof}
It can be verified that $\Phi_k$ is the extrinsic MAP
estimator of $X_k$. We have
\begin{align*}
\gmap_k(\alpha) &= \frac{\partial H(\X|\Y(\alpha_1,\cdots,\alpha_n))}{\partial\alpha_k} = \frac{\partial H(X_k|Y_k,\Phi_k)}{\partial \alpha_k}.
\end{align*}
For notational convenience, we omit the dependence of $X_k$, $Y_k$ and $\Phi_k$ on the symbol index $k$ whenever possible. The conditional entropy of $X_k$ is given by
\begin{align*}
  &H(X|Y,\Phi)\! =\! -\int\limits_{y,\phi}\!\!\sum_{x\in\Xc}p(x,y,\phi)\log_2\frac{p(x,y,\phi)}{\sum_{x'\in\Xc}p(x',y,\phi)}\text{d}y\text{d}\phi\\
&=\!\!
\sum_{x\in\Xc}\!p(x)\!\!\!\int\limits_{\phi}\!\!\!p(\phi|x)\!\!\left(\!\int\limits_y\!\!\!p(y|x)\!\log_2\!\frac{\sum_{x'\in\Xc}p(x'|\phi)p(y|x')}{p(x|\phi)p(y|x)}\text{d}y\!\!\right)\!\!\text{d}\phi.
\end{align*}
This follows by noting that $p(x_k,y_k,\phi_k) = p(y_k|x_k)p(\phi_k|x_k)p(x_k)$. The result now follows by noting that $p(x_k|\phi_k) = p(x_k|\y_{\sim k})$.

Assuming that each bit in the symbol $x_k$ is independent (which is true
asymptotically as $N\to\infty$), we have
\begin{align}
\label{eq:utov}
\u\! &=\! \left\{\!\prod_{m=1}^Mp(x_k^{[m]}|\y_{\sim k}),x_k\in\Xc\!\right\}\notag \\
&\triangleq \{f_{x_k}(v_1,\cdots,v_M),x_k\in\Xc\},
\end{align}
where $v_m = \log\frac{p(x_k^{[m]}=+1|\y_{\sim
    k})}{p(x_k^{[m]}=-1|\y_{\sim k})}$. From \eqref{eq:utov}, we see
that $\u$ is completely characterized by $v_1,\cdots,v_M$. So,
\eqref{eq:kernel_bicm} is henceforth interpreted in terms of the
log-likelihood ratios $v_m$. 
\end{IEEEproof}
% The asymptotic GEXIT function is given by 
% \begin{align*}
% %  \label{eq:agexit}
%   \gmap(\alpha) = \limsup_{N\to\infty}\mathbb{E}\left[\frac{1}{N}\sum_{k=1}^{N/M}\gmap_k(\alpha)\right].
% \end{align*}

The density $\x_{k}(\u)$ in~\eqref{eq:lgexit} is hard to compute, so
instead one can use the BP estimate to compute the density for the
asymptotic limit $N\to\infty$. The curve obtained by using the BP
estimate is called the BP-GEXIT function $\gbp(\alpha)$. Let
$\x^{(\ell)}(v)$ denote the density of the log-likelihood ratio,
conditioned on the transmission of the all-zero codeword, emitted from
the variable nodes to the detector nodes, during iteration $\ell$. For
a fixed $\ell$, as the blocklength $N\to\infty$, we have $p_x(v_m) =
\x^{(\ell)}\left(x^{[m]}v_m\right)$. If $F[\x]$ is the
density transformation operator corresponding to the map $\mu\mapsto
\text{e}^{\mu}/(1+\text{e}^{\mu})$, then we can write
\begin{align}
  \label{eq:udensity}
  \x(\u) &= \{p(\u|x),x\in\Xs\}\notag \\
  &= \left\{\prod_{m=1}^MF[\x](x^{[m]}v_m),x\in\Xs\right\},
\end{align}
where $\u$ is given by \eqref{eq:utov}. The BP-GEXIT function $\gbp$ is
computed as follows: For a given channel parameter $\alpha$, compute
the fixed point of density evolution, say $\a$. Then,
\begin{align*}
  \gbp(\alpha) = \gfun(F[L(\rho(\a))];\alpha).
\end{align*}

One can now calculate an upper bound on the MAP decoding
threshold. The following procedure is now fairly standard and the
details can be found in~\cite[Sec. 4.12]{RU-2008}. It can be shown
that the GEXIT functional $\gfun$ preserves degradation. So,
by the optimality of the MAP decoder, the GEXIT function always lies
below the BP-GEXIT function i.e., $\gmap(\alpha)\leq\gbp(\alpha)$. Let
$\amapub$ be the largest positive number such that
\begin{align*}
  \int_{\amapub}^1\gbp(\alpha)\text{d}\alpha = \rate(\lambda,\rho).
\end{align*}
From the properties of the BP-GEXIT function and the GEXIT function,
we have $\amap\leq\amapub$. Following~\cite{Kudekar-isit12}, we refer
to $\amapub$ as the area threshold.

In this work we compute the BP-GEXIT curves for the case when there is
no demapper update between iterations. The BP-GEXIT functions and the
area threshold, for different BICM schemes are shown in
Figures~\ref{fig:gexit_qpsk},~\ref{fig:gexit_16qammap},~\ref{fig:gexit_16qammlm}. The
parameter $\alpha$ was chosen to be the normalized channel entropy per
bit.
%%% Local Variables: 
%%% mode: latex
%%% TeX-master: "bicm_paper"
%%% End: 

\section{Spatially-Coupled LDPC Codes}
\label{sec:spat-coupl-ldpc}
In this section, we describe the \sc{} $(\dl,\dr,L,w)$ ensemble
introduced in~\cite{Kudekar-it11}. The variable nodes are placed at
positions $[-L,L]$ and the check nodes are placed at positions
$[-L,L+w-1]$ ($w$ can be thought of as a ``smoothing''
parameter). Each of the $\dl$ connections, of a variable node at
position $i$, are uniformly and independently chosen from $[i,i+w-1]$
as shown in Fig.~\ref{fig:sc_code}. There are two consequences of
this coupling - threshold increase and rate
loss~\cite{Kudekar-it11}. The design rate of a \sc{} $(\dl,\dr,L,w)$
ensemble is given by
\begin{align*}
  \rate(\dl,\dr,L,w) = \left(1 - \frac{\dl}{\dr}\right) - \frac{\dl}{\dr}\frac{w+1-2\sum_{i=0}^w\left(\frac iw\right)^{\dr}}{2L+1}.
\end{align*}
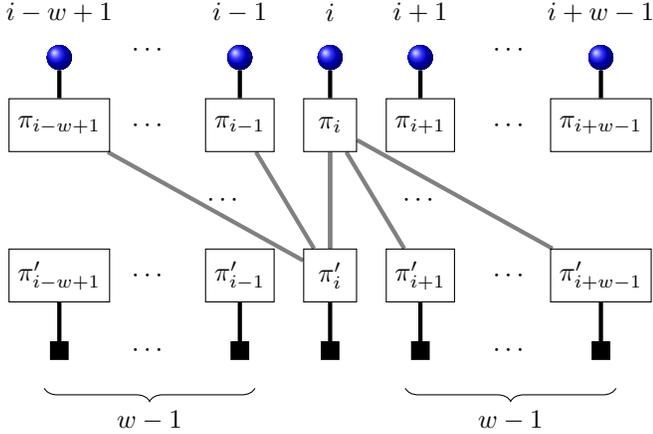
\begin{figure}
  \centering
  \begin{tikzpicture}[xscale=0.4,yscale=1]
  \foreach \i/\text in {2/i-w+1,8/i-1,11/i,14/i+1,20/i+w-1} { 
    \node[ball color=blue,circle]
    (f_node\i) at (\i,3.9) {}; 
    \node[draw=black,fill=black,rectangle] (g_node\i) at (\i,0)
    {};
    \node[draw,rectangle,minimum size=20pt] (p1_node\i) at (\i,3) {$\pi_{\text}$};
    \node[draw,rectangle,minimum size=20pt] (p2_node\i) at (\i,1)
    {$\pi_{\text}'$};
    \draw[ultra thick] (f_node\i) -- (p1_node\i);
    \draw[ultra thick] (g_node\i) -- (p2_node\i);
} 
\draw[decorate,decoration={brace,amplitude=5pt}]
  (8.5,-0.5) -- node[midway,below=5pt] {$w-1$} (1.5,-0.5);
\draw[decorate,decoration={brace,amplitude=5pt}] (20.5,-0.5) --
  node[midway,below=5pt] {$w-1$} (13.5,-0.5); 

\foreach \i/\text in {2/i-w+1,8/i-1,11/i,14/i+1,20/i+w-1} 
 \node at (\i,4.5) {$\text$};
 \begin{pgfonlayer}{background}
\foreach \i/\j in {11/2,11/8,11/11} {
\draw[ultra thick,gray] (p2_node\i) -- (p1_node\j);
}
\foreach \i/\j in {11/11,11/14,11/20} {
\draw[ultra thick,gray] (p1_node\i) -- (p2_node\j);
}
\node (tmp) at (7.5,2) {$\cdots$};
\node (tmp) at (14,2) {$\cdots$};
 \end{pgfonlayer}
\foreach \x in {5,17} {
\foreach \y in {4,3,1,0} { 
\node (blu) at (\x,\y) {$\cdots$};
}}
\end{tikzpicture}

%%% Local Variables: 
%%% mode: latex
%%% TeX-master: "../bicm_paper"
%%% End: 
  \caption{A portion of a generic spatially-coupled code. The variable
    nodes are shown as blue filled circles and the check nodes are
    respresented by black squares. The variable node at position $i$
    is connected to check nodes at positions $[i,i+w-1]$. The
    permutations $\pi_i$ are chosen uniformly at random.}
  \label{fig:sc_code}
\end{figure}
Let $\a_i^{(\ell)}$ be the average density, in the spirit
of \eqref{eq:avg_density}, emitted by the variable nodes at position
$i$. Set $\a_i^{(\ell)} = \Delta_{+\infty}$, the Dirac delta
function at infinity, for $i\notin[-L,L]$. The DE for this ensemble
can be written as
\begin{align}
\label{eq:sc_de_bicm}
  \a_i^{(\ell+1)} &=
  \phi\left(L(\x^{(\ell)}_i);\sigma\right)\varoast\lambda(\x^{(\ell)}_i)\notag \\
\x^{(\ell)}_i &= \frac 1w\sum_{j=0}^{w-1}\rho\left(\frac 1w\sum_{k=0}^{w-1}\a_{i+j-k}^{(\ell)}\right).
\end{align}
Here $L(\x) = \x^{\varoast\dl}, \lambda(\x) = \x^{\varoast\dl-1},$ and
$\rho(\x) = \x^{\boxast\dr-1}$. For a channel parameter $\alpha$, let
$\bar{\a} = [\a_{-L},\cdots,\a_L]$ denote the fixed point implied by
\eqref{eq:sc_de_bicm}. Using the technique developed in
Section~\ref{sec:gexit-curves}, one can compute the GEXIT curves for
\sc{} LDPC codes from (\ref{eq:lgexit}) along the lines of
\cite{Kudekar-it11}. Define the GEXIT functional for \sc{} codes via
\begin{align*}
  \gfun(\bar{\a}) \triangleq \frac{1}{2L+1}\sum_{i=-L}^L\gfun(\a_i),
\end{align*}
where $\gfun(\a)$ is defined in \eqref{eq:lgexit}. The BP-GEXIT
function of \sc{} codes is given in parametric form by
$(\alpha,\gfun(\bar{\a}))$.

%%% Local Variables: 
%%% mode: latex
%%% TeX-master: "bicm_paper"
%%% End: 

\section{Results}
\label{sec:results}
\begin{figure}
  \centering
  \input{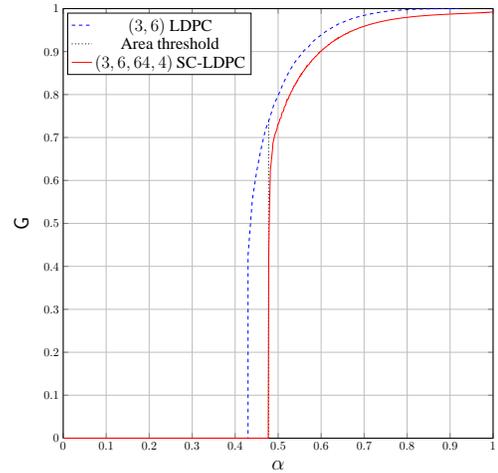}
  \caption{The BP-GEXIT curve of a $(3,6)$ code for QPSK modulation and the upper bound on the MAP  threshold computed via the area theorem. Also shown is the BP-GEXIT curve of the $(3,6,32,4)$ \sc{} ensemble.}
  \label{fig:gexit_qpsk}
\end{figure}

We consider the case when there is no demapper update between
iterations. The BP-GEXIT curve of a $(3,6,64,4)$ spatially-coupled
ensemble is shown in
Figures~\ref{fig:gexit_qpsk},~\ref{fig:gexit_16qammap},~\ref{fig:gexit_16qammlm}
for different BICM
schemes. Figures~\ref{fig:gexit_qpsk},~\ref{fig:gexit_16qammap},~\ref{fig:gexit_16qammlm}
also demonstrate the threshold saturation phenomenon, wherein the BP
threshold of the spatially-coupled ensemble is close to an intrinsic
threshold of the underlying ensemble. For the optimal MAP demapper, this
intrinsic threshold is the area threshold computed using the BP-GEXIT
curve of the underlying ensemble. From Fig.~\ref{fig:gexit_16qammlm},
we observe that for the suboptimal MLM demapper, the BP threshold of
the \sc{} code crosses the area threshold of the underlying
ensemble.

It is known that increasing the left degree of regular LDPC codes,
while keeping the rate constant, pushes the MAP threshold of the
ensemble towards the noise threshold for BMS channels. Based on this,
we compute the BP thresholds of a $(4,8,64,4)$ and $(6,12,64,4)$
spatially-coupled ensemble and as seen from
Tables~\ref{tab:bicm_universality_map} and
~\ref{tab:bicm_universality_mlm}, the gap to the BICM noise threshold
indeed becomes smaller. The BP thresholds and the noise thresholds for
the Rayleigh fast fading channel are shown in
Tables~\ref{tab:bicm_universality_map}
and~\ref{tab:bicm_universality_mlm}. The tables also show the
asymptotic gap as the rate loss tends to zero, i.e., when
$L\to\infty$. Based on this numerical evidence, we conjecture that the
ensemble of spatially-coupled LDPC codes with large left degrees
universally approach the noise threshold for different BICM schemes.

The decoding thresholds of coding schemes for BICM can be improved by
using BICM-ID. For comparison, the thresholds of the $(4,8,64,4)$
\sc{} ensemble is computed for BICM-ID. The thresholds are computed
for $64$QAM modulation using a MAP demapper and we consider
transmission over both AWGN and Rayleigh fast-fading channels. In this
case, the demapper update was performed once for every $100$
iterations of the \sc{} code. As expected, the gap to the noise
threshold reduces to $0.09$ dB from $0.20$ dB and $0.24$ dB for AWGN
and Rayleigh fast-fading channels respectively. This improved
performance comes at the expense of increased complexity.

\begin{figure}
  \centering
  \input{./figures/gexit_36324_16qammap}
  \caption{The BP-GEXIT curve of a $(3,6)$ code for $16$QAM modulation with a MAP detector and the upper bound on the MAP  threshold computed via the area theorem. Also shown is the BP-GEXIT curve of the $(3,6,32,4)$ \sc{} ensemble.}
  \label{fig:gexit_16qammap}
\end{figure}

\begin{figure}
  \centering
  \input{./figures/gexit_36324_16qammlm}
  \caption{The BP-GEXIT curve of a $(3,6)$ code for $16$QAM modulation with an MLM detector and the upper bound on the MAP  threshold computed via the area theorem. Also shown is the BP-GEXIT curve of the $(3,6,32,4)$ \sc{} ensemble.}
  \label{fig:gexit_16qammlm}
\end{figure}

\begin{table*}[t!]\hfill{}
  \caption{Performance of various \sc{} ensembles over BICM channels with a MAP demapper.}
  \label{tab:bicm_universality_map}
  \centering
%  \small
  \makebox[\linewidth]{
  \begin{tabular}{|c|c|c|c|c|}
    \cline{3-5}
    \multicolumn{2}{c}{} & \multicolumn{3}{|c|}{BP Thresh.\,/\,Gap\,/\,Asympt. Gap ($E_b/N_0$ in dB)} \\ \hline
    Mod.  / Chan.  & Noise Thresh. ($E_b/N_0$ in dB) & $(3,6,64,4)$ & $(4,8,64,4)$ & $(6,12,64,4)$ \\ \cline{1-5}
    QPSK  / AWGN   & 0.17                            & 0.57\,/\,0.40\,/\,0.31 & 0.54\,/\,0.37\,/\,0.11 & 0.33\,/\,0.16\,/\,0.06 \\ \hline
    16QAM / AWGN   & 2.27                            & 2.71\,/\,0.44\,/\,0.35 & 2.49\,/\,0.22\,/\,0.13 & 2.43\,/\,0.16\,/\,0.07 \\ \hline
    64QAM / AWGN   & 4.67                            & 5.11\,/\,0.44\,/\,0.35 & 4.87\,/\,0.20\,/\,0.11 & 4.82\,/\,0.15\,/\,0.05 \\ \hline
    QPSK  / Fading & 1.83                            & 2.27\,/\,0.44\,/\,0.35 & 2.04\,/\,0.21\,/\,0.12 & 2.00\,/\,0.17\,/\,0.07 \\ \hline
    16QAM / Fading & 4.11                            & 4.56\,/\,0.45\,/\,0.36 & 4.34\,/\,0.23\,/\,0.14 & 4.29\,/\,0.18\,/\,0.08 \\ \hline
    64QAM / Fading & 6.62                            & 7.11\,/\,0.49\,/\,0.40 & 6.86\,/\,0.24\,/\,0.15 & 6.80\,/\,0.18\,/\,0.08 \\ \hline
  \end{tabular}
}
\hfill{}
\end{table*}
\begin{table*}[t!]\hfill{}
  \caption{Performance of various \sc{} ensembles over BICM channels with an MLM demapper.}
  \label{tab:bicm_universality_mlm}
  \centering
%  \small
  \makebox[\linewidth]{
  \begin{tabular}{|c|c|c|c|c|}
    \cline{3-5}
    \multicolumn{2}{c}{} & \multicolumn{3}{|c|}{BP Thresh.\,/\,Gap\,/\,Asympt. Gap ($E_b/N_0$ in dB)} \\ \hline
    Mod.  / Chan.  & Noise Thresh. ($E_b/N_0$ in dB) & $(3,6,64,4)$ & $(4,8,64,4)$ & $(6,12,64,4)$ \\ \cline{1-5}
    QPSK  / AWGN   & 0.17                            & 0.57\,/\,0.40\,/\,0.31 & 0.54\,/\,0.37\,/\,0.11 & 0.33\,/\,0.16\,/\,0.06 \\ \hline
    16QAM / AWGN   & 2.29                            & 2.70\,/\,0.41\,/\,0.32 & 2.51\,/\,0.22\,/\,0.13 & 2.47\,/\,0.18\,/\,0.08 \\ \hline
    64QAM / AWGN   & 4.71                            & 5.14\,/\,0.43\,/\,0.34 & 5.00\,/\,0.29\,/\,0.20 & 4.96\,/\,0.25\,/\,0.15 \\ \hline
    QPSK  / Fading & 1.83                            & 2.27\,/\,0.44\,/\,0.35 & 2.04\,/\,0.21\,/\,0.12 & 2.00\,/\,0.17\,/\,0.07 \\ \hline
    16QAM / Fading & 4.17                            & 4.63\,/\,0.46\,/\,0.37 & 4.41\,/\,0.24\,/\,0.15 & 4.36\,/\,0.19\,/\,0.09 \\ \hline
    64QAM / Fading & 6.73                            & 7.26\,/\,0.53\,/\,0.44 & 7.01\,/\,0.28\,/\,0.19 & 6.95\,/\,0.22\,/\,0.12 \\ \hline
  \end{tabular}
}
\hfill{}
\end{table*}
%}
%\end{center}

%%% Local Variables: 
%%% mode: latex
%%% TeX-master: "bicm_paper"
%%% End: 

\section{Concluding Remarks}
\label{sec:concluding-remarks}
Spatially-coupled LDPC codes have shown promising results for a large
class of graphical models. In this work, we study their performance on
BICM channels and validate the conjecture that the phenomenon of
threshold saturation is indeed very general. We extend the GEXIT
analysis of LDPC codes for BICM schemes. This enables one to bound the
performance of the MAP decoder of LDPC codes. The area threshold,
which upper-bounds the MAP decoding performance, of LDPC codes is
computed for different demappers and modulations. Using these tools,
we numerically demonstrate the phenomenon of threshold saturation for
these channels. We note that when using suboptimal demappers (like the
MLM demapper), the threshold saturates towards an intrinsic threshold
of the suboptimal demapper and the upper bound computed using GEXIT
curves is no longer tight. This is consistent with previous results
for spatially-coupled systems with suboptimal component
decoders~\cite{Takeuchi-isit11,Schlegel-isit11,Jian-isit12}. The
performance also improves significantly when used with BICM-ID and the
thresholds of SC-LDPC codes approach the noise threshold with smaller
degrees. These asymptotic results demonstrate that SC-LDPC codes
approach the noise threshold of different BICM schemes.

Irregular LDPC codes and multi-edge type LDPC codes were designed for
BICM schemes in \cite{Hou-it03,Sankar-globe04} with excellent
thresholds. In~\cite{Nguyen-globe11}, the authors design families of
protograph-based LDPC codes for BICM schemes with excellent thresholds
of around $0.2$-$0.4$ dB from the BICM noise threshold. For finite
$L$, spatially-coupled codes compare favorably with the previously
optimized LDPC ensembles in-spite of the rate loss incurred due
to finite $L$.

%%% Local Variables: 
%%% mode: latex
%%% TeX-master: "bicm_paper"
%%% End: 

%use section* for acknowledgement
%\section*{Acknowledgment}


\ifCLASSOPTIONcaptionsoff
  \newpage
\fi



%\bibliographystyle{IEEEtran}
\bibliographystyle{../../slpbib/IEEEtran}
\bibliography{../../slpbib/IEEEabrv,../../slpbib/IEEEfull,../../slpbib/WCLfull,../../slpbib/WCLabrv,../../slpbib/WCLnewbib,../../slpbib/WCLbib}

%\begin{biography}[{\includegraphics[width=1in,height=1.25in,clip,keepaspectratio]{mshell}}]{Michael Shell}

% \begin{IEEEbiographynophoto}{Arvind Yedla}
% \end{IEEEbiography}

% \begin{IEEEbiographynophoto}{Mostafa El-Khamy}

% % insert where needed to balance the two columns on the last page with
% %\newpage

% \begin{IEEEbiographynophoto}{Jungwon Lee}

% \begin{IEEEbiographynophoto}{Inyup Kang}

%\vfill


\end{document}